\documentclass[showpacs,superscriptaddress,preprintnumbers]{revtex4}
\usepackage{amsfonts}
\usepackage{amssymb}
\usepackage{amsmath}
\usepackage{epsfig}
\usepackage{float}
\usepackage{subfig}
\usepackage{array}
\usepackage{xcolor}
\usepackage{mathrsfs}
\allowdisplaybreaks
\begin{document}
\title{Charged Fuzzy Dark Matter Black Holes}

\author{Z. Yousaf}
\email{zeeshan.math@pu.edu.pk}
\altaffiliation{Corresponding author}
\affiliation{Department of Mathematics, University of the Punjab, Lahore-54590, Pakistan.}

\author{Bander Almutairi}
\email{baalmutairi@ksu.edu.sa}
\affiliation{Department of Mathematics, College of Science, King Saud University, P.O.Box 2455 Riyadh 11451, Saudi Arabia}

\author{S. Khan}
\email{suraj.pu.edu.pk@gmail.com}
\affiliation{Department of Mathematics, University of the Punjab, Lahore-54590, Pakistan.}

\author{Kazuharu Bamba}
\email{bamba@sss.fukushima-u.ac.jp}
\affiliation{Faculty of Symbiotic Systems Science,
Fukushima University, Fukushima 960-1296, Japan}


\begin{abstract}

We investigate the impact of fuzzy dark matter (FDM) on supermassive black holes (SMBHs) characterized by a spherical charge distribution. This work introduces a new class of spherically symmetric, self-gravitational relativistic charged models for FDM haloes, using the Einasto density model. This study enables the dark matter (DM) to appear as the matter ingredient, which constructs the black hole and extends the non-commutative mini black hole stellar solutions. By considering the charged anisotropic energy-momentum tensor with an equation of state (EoS) $p_{r}=-\rho$, we explore various black hole solutions for different values of the Einasto index and mass parameter. Our approach suggests that the central density of the resulting black hole model mimics the usual de Sitter core. Furthermore, we discuss the possibility of constructing a charged self-gravitational droplet by replacing the above-mentioned EoS with a non-local one. However, under these circumstances, the radial pressure is observed to be negative. Ultimately, we consider various possibilities of constructing DM black holes, featuring intermediate masses that could evolve into galaxies. Consequently, some of these theoretical models have the potential to replace the usual black hole solutions of the galactic core. Simultaneously, these models are physically beneficial for being comprised of the fundamental matter component of the cosmos. Due to the outcomes of this paper, we would be able to study the connection between BH and DM by formulating stable stellar structures featuring fuzzy mass distributions derived from the Einasto distribution of DM halos.

\end{abstract}
\maketitle

\section{Introduction}

The relationship between astrophysical dark entities, such as black
holes (BHs) and dark matter (DM), presents a fascinating and complex
challenge in modern cosmology and astrophysics. Compact objects like
BHs are cosmic vacuums where the gravitational field is highly
dominant, opening new avenues for understanding the interaction
between gravity and fundamental physics
\cite{brown2016complexity,alishahiha2019black,yousaf2024role}. They
originate from the collapse of massive stars or the merging of dense
objects and have been thoroughly studied using Einstein's
gravitational model. On the other hand, DM makes up about $27\%$ of
the total mass energy of the universe and remains an intriguing
puzzle for researchers \cite{arkani2009theory,yousaf2024imprints}.
Despite its pivotal influence in shaping large-scale cosmic
structures, DM has not yet been directly observed. Different models
have been suggested to explain the nature of DM, with the fuzzy dark
matter (FDM) model emerging as a strong contender. The FDM model has
attracted the attention of researchers due to its unique
characteristics and implications for cosmic structures.

The FDM model is defined by its wave-like features, which stem from
the quantum mechanical nature of light bosons with masses around
$10^{-22}$ eV, including axions and other similar particles. The
wave-like nature of FDM distinguishes it from other DM candidates,
like cold DM, and has significant consequences for the structure and
development of self-gravitating cosmic objects. Unique cosmic
phenomena, such as interference patterns in density distributions,
smooth cores in dark matter halos, and inhibited structure formation
at small scales, are observed in FDM at both galactic and cosmic
scales. These unique properties of FDM offer a fresh perspective on
DM, particularly in regions of strong gravitational fields, like
those near BHs. The motivation for exploring FDM in the
context of BHs stems from the crucial role that DM, in any of its
forms, plays in the dynamics of self-gravitational stellar systems.
The existence of DM in close vicinity to BHs can affect their
stability, evolution, and observable characteristics, including
emission spectra, accretion processes, and gravitational wave
signals. The distinctive quantum features associated with FDM
further intensify its interaction with BHs, especially in regions of
high density and intense curvature. The dynamics of matter and
energy surrounding BHs can be affected by electromagnetic fields,
which introduces another dimension to the study of BHs
\cite{jimenez2021charged}. Electromagnetic radiation can interact
with BHs, especially those that are charged or surrounded by
magnetic fields
\cite{vogelsberger2012subhaloes,zavala2013constraining,vogelsberger2013direct}.
A wide range of phenomena are produced by this interaction,
including the production of relativistic jets, dynamics of the
accretion disk, and energy extraction through the Penrose process.
Under these conditions, FDM may alter the structure of the
electromagnetic field and affect its behavior in ways that classical
DM theories do not foresee
\cite{ansoldi2007non,nicolini2006noncommutative}. Studying this
interaction is vital for understanding the observation data from BH
systems. This data includes X-ray emissions, radio signals, and the
recently detected gravitational wave signals resulting from BH
mergers.

{A key reason for exploring the fuzzy FDM model is
its quantum mechanical properties. The existence of ultra-light
bosonic particles in the FDM model produce wave-like
characteristics on galactic scales. The ``cusp-core'' and ``missing
satellite'' issues that are frequently encountered in cold DM
simulations could be resolved by the FDM model by naturally
smoothing out small-scale density fluctuations. Although
observations have limited the FDM model, it remains a
viable possibility. Certain astronomical data, such as the matter
distribution in galactic halos and the rotation curves of dwarf
galaxies, indicate that FDM might still be a good fit in some
regimes. Based on the particular model under consideration, the
limitations of the FDM model could alter when more accurate data
become available. Other alternatives, such as mixed
DM models or more complex FDM interactions, are still being
explored, even if existing observations may rule out some parameter
ranges (such as certain masses of FDM particles). There is still
much to learn about the whole effect of FDM on the creation of
stellar structures. Investigating the impact of ultralight scalar
fields on gravitationally bound stellar structures could open up
new directions in fundamental physics, including possibly relating
DM to inflationary models in the early universe or string
theory.}

To acquire a better knowledge of the nature of BH, it is essential
to explore their interior composition in more detail. This kind of
study can be conducted from a geometric perspective
\cite{borde1997regular,brady1999internal,nomura2015black}.
Furthermore, understanding the type of matter that contributed to
the BH's formation is equally essential for a comprehensive
understanding \cite{bambi2013non,malafarina2015compact}. Since
baryonic and leptonic matter makes up only $4\%$ of the entire cosmic
content, the density profiles of DM increase as we approach the
galactic center, highlighting the importance of exploring the
interplay between DM and BHs
\cite{horowitz2004black,koch2005black,ahn2008black}. Since baryonic
and leptonic matter makes up only $4\%$ of the entire cosmic content,
the density profiles of DM increase as we approach the galactic
center, highlighting the importance of exploring the interplay
between DM and BHs. The no-hair theorem generally prevents us from
observing the interior attributes of BHs
\cite{ruffini1971introducing,heusler1996black,gurlebeck2015no}.
However, modeling the interior configuration of a BH, which is an
established technique in the literature
\cite{sakharov1966initial,dymnikova1992vacuum,dymnikova2002cosmological},
has the potential to show whether a relationship between DM and BHs
is possible. A recent effort in this direction was presented in
\cite{yousaf2024generating,khan2024structure,khan2024construction},
where the researchers model the central galactic object by assuming
a DM profile that is fitted to the outer regions of the galaxy.
Other potential relationships between the two key components of the galactic bulge were explored in the studies \cite{boshkayev2019model,boshkayev2019model} (and references
therein), which further supports the idea that the BH plays a
central role in galactic structure. Based on this scenario, we
developed a new model of the central galactic object as a fuzzy BH
or a self-gravitating compact droplet structure. This model is
closely related to BH/droplet models inspired by non-commutative
geometry \cite{nicolini2009noncommutative}, which feature a Gaussian
matter distribution and a de Sitter equation of state (EoS). To
successfully accomplish our objectives, we will couple the proposed
density profile with a stress-energy tensor representing an
anisotropic fluid and an EoS of the type $p_{r}= -\rho$. This form
of EoS appears frequently in investigations of BHs
\cite{dymnikova1992vacuum,dymnikova2002cosmological,batic2021fuzzy,batic2022possible}.
In this case, it produces different kinds of regular BH models.
Before the possible existence of a supermassive BH at the heart of
the Milky Way, known as Sagittarius $A^{\ast}$, various theoretical
models were developed that replaced the central BH with other
gravitational objects. Gravastars \cite{mazur2004gravitational},
boson stars \cite{ruffini1969systems}, naked singularities
\cite{chowdhury2012circular}, burning disks
\cite{chowdhury2012circular}, quantum cores \cite{ruffini2015core},
and gravitationally bound clumps of DM based on the
exponential-sphere density
parameterizations \cite{levkov2018gravitational} featured among
them. Apart from the aforementioned endeavors, numerous other
researchers have explored alternative mechanisms that could initiate
the formation of SMBHs in the galactic centers. For example, the
formation of traversable wormholes within the outer galactic halos
based on Einasto parameterizations have been discussed in
\cite{ovgun2016existence}. Also, the formation of DM wormholes using
the NFW and King's density profiles have been examined in
\cite{islam2019formation}, while \cite{ovgun2021evolving}
investigated the evolution of topologically deformed DM wormholes
subject to the Chaplygin gas EoS with different density functions.
By using three different DM density functions, the influence of DM on
the weak deflection angle by central galactic BHs has been analyzed
in \cite{pantig2022dark}. More recently, the authors of
\cite{ovgun2023constraints} constructed four different spherically
symmetric BH models immersed into DM halos using the Generalized
Uncertainty Principle, while the possible formation of BHs with DM
halos  dwarf galaxies is presented in \cite{pantig2022dehnen}.
Furthermore, the effect of DM on the quasinormal modes and
quasibound states of SMBHs have been probed in
\cite{liu2023gravitational}. The effect of FDM on SMBHs using
spherical matter distribution has been explored in
\cite{pantig2023black}. They showed that FDM establishes a soliton
core that surrounds the SMBH at the galactic center. The spherically
symmetric BHs based on pseudo-isothermal DM profile have been
derived in \cite{yang2023black}.

On the other hand, the Einasto density model serves as a
mathematical model employed to investigate the distribution of
matter, including DM halos, around black holes. Retana-Montenegro et
al. \cite{retana2012analytical} studied different analytical
features associated with the three-parameter Einasto density model and
discussed the relevance of this DM halos model in comparison to
other density profiles. Numerous scientists have developed various
black hole solutions by employing the Einasto density profile.
Motivated by the prevalence of DM in galactic cores, Batic et al.
\cite{batic2021fuzzy} explored the effects of FDM on the
supermassive stellar structures existing within the galactic
structures. They investigated the possible formation of
self-gravitational structures based on the FDM model, utilizing an
anisotropic fluid distribution based on the Einasto density
function. In their subsequent work \cite{batic2022possible}, the
same authors examined the feasibility of stable self-gravitational
objects with fuzzy mass distribution motivated by standard DM
density functions. These astronomical entities manifest in three
forms: stable fuzzy self-gravitational droplets (non-horizon), and
fuzzy BHs characterized by either one or two horizons. Figueiredo et
al. \cite{figueiredo2023black} developed an asymptotically flat BH
metric with anisotropic matter configuration using generic density
models. These models exhibit several realistic astrophysical
scenarios and can describe the galactic structures hosting SMBHs
surrounding DM. Baes \cite{baes2022einasto} presented a systematic
approach to explore the dynamical and photometric structure of the
range of Einasto models across the complete spectrum of model
parameter space, in the context of DM halos.

This manuscript seeks to explore the combined effects of FDM and the
electromagnetic field on BHs, focusing on their impact on the BH's
properties and the surrounding spacetime geometry. Taking into
account the wave-like character of the FDM model, we explore its
quantum interference patterns and localized wave structures that interact
with the gravitational and electromagnetic fields of the BH in the
background of the Einasto density profile. These interactions are
believed to impact important BH features such as accretion rates and
energy output, which might be detected by astronomical data. The new
features identified in this work might shed light on the mechanics
of BHs and DM, which could lead to new possibilities for the
indirect detection of FDM through astronomical measurements. The
structure of this study is as follows: In the coming section, we
define the Einasto density model and some associated definitions
such as the mass function, lower incomplete Gamma function, and the
density profile. In Sec. \textbf{III} and \textbf{IV}, we establish
charged FDM structures subject to the de Sitter-type EoS, the
corresponding effective potential their viability of regenerating
the dynamics of stellar systems. Section \textbf{V} is devoted to
constructing the self-gravitating fuzzy-charged DM composed of
anisotropic fluid for the stable stats of massive particles, under
the assumption of non-local EoS. Consequently, the conclusion follows
at the end of the manuscript.

\section{The Einasto Density Model}

In 1965, Einasto \cite{einasto1965kinematics} presented an alternative density model, which is generally used to study the distribution of cold DM halos in galaxy clusters.
Einasto \cite{einasto1969galactic,einasto1969andromeda} showed that the gravitational potential, mass density, and the cumulative mass profile are particular types of descriptive
functions that are involved in the realistic modeling of the galactic system. These descriptive functions provide a faithful characterization of the system and
are the fundamentals of the density function $\rho(r)$. Therefore, it is reasonable to assume the density configuration itself as the primary and fundamental descriptive
function of a galactic model. This configuration should  manifest the following characteristics
\begin{itemize}
\item Jump discontinuities should not be present in the above-mentioned descriptive functions.
\item The density function $\rho(r)$ is a smooth and asymptotically zero. That is, $\rho(r)\in C^{\infty}(\mathbb{R}^{+})$ with \begin{align}\nonumber
\lim_{r\rightarrow\infty}\rho(r)=0.
\end{align}
\item $0<\rho(r)<\infty$
~$\forall ~r>0$.
\item  The system's total mass, effective radius, and central gravitational potential, all of which are linked to $\rho$, should have finite values.
\end{itemize}
Besides modeling different galaxies such as Sculptor dwarfs, M31, M32, M87, and Milky Way \cite{einasto1969galactic,einasto1969andromeda}, the Einasto DM model can also be used in describing the density of DM haloes \cite{navarro2004inner,springel2005simulations,mamon2005dark,
hayashi2008understanding,gao2008redshift}. Further, some analytical studies inspired by the Einasto DM model include the spherically symmetric galaxy (spiral and analytical) models, and their DM haloes characterized by the logarithmic slope have also been discussed in the literature \cite{cardone2005spherical,dhar2010surface,retana2012analytical}. Since the Einsato density model is generally considered in the simulation of $\Lambda$ cold DM haloes \cite{gao2008redshift,hayashi2008understanding,merritt2006empirical}, local DM density using galactic velocity curve \cite{de2019estimation}, structural features of spiral as well as eliptical galaxies, bulges and bars using SDSS survey \cite{gadotti2009structural} and the properties of dwarf elliptical galaxies \cite{graham2003hst}. The Einasto model \cite{einasto1969galactic,einasto1969andromeda} is characterized by the density
function
\begin{align}\label{s1}
\rho(r)=\rho_{s}\exp\left\{-d_{\beta}\left[\left(\frac{r}{r_{s}}\right)^\frac{1}{\beta}-1\right]\right\},
\end{align}
where $r_{s}$ is the half-mass radius, $d_{\beta}$ is the numerical
constant controlling $r_{s}$, $\rho_{s}$ denotes the central density at
$r=r_{s}$ and $\beta$ is the Einasto index.
The Einasto density model can be parameterized in several ways as
can be seen in the literature. However, one popular representation
of the model for DM halos is defined by
\begin{align}\label{s2}
\rho(r)=\rho_{-2}\exp\left\{-2\beta\left[\left(\frac{r}{r_{-2}}\right)^\frac{1}{\beta}-1\right]\right\}.
\end{align}
Here, $r_{-2}$ and $\rho_{-2}$ denote the radius and density at
which $\frac{d\ln \rho}{d\ln r}=-2$. Now, by assuming the scale
length
\begin{align}\label{s3}
h=\frac{r_{-2}}{(2\beta)^\beta}=\frac{r_{s}}{d^\beta_{\beta}},
\end{align}
and the value of the charged central density
\begin{align}\label{s4}
\rho_{0}+e_{0}=(\rho_{s}+e_{s})\exp(d_{\beta})=\rho_{-2}\exp(2\beta),
\end{align}
where $e_{s}$ denotes the central charge.
{Furthermore, $\rho_{0}$ and $e_{0}$ denote the
values of density and charge at $r=0$.} Using this value in Eq.
\eqref{s2}, we get
\begin{align}\label{s5}
\rho(r)=(\rho_{0}+e_{0})\exp\left[-\left(\frac{r}{h}\right)^{\frac{1}{\beta}}\right],
\end{align}
The charged DM density profile based on Einasto's parameterization consists of numerous components, each of which is dependent on its own four sets of parameters $\{\rho_{0},\beta,h, e_0\}$. Therefore, it is possible to simulate a variety of stellar structures by giving certain values of parameters. For example, in the case of DM haloes with masses in the range of dwarf galaxies to rich galaxy clusters, $4.54\lesssim\beta\lesssim8.33$, where the average value of $\beta=5.88$ \cite{navarro2004inner}. It has been analyzed that the value of $\beta$ decreases with redshift and mass, with $\beta\sim4.35$ in the case of cluster-sized and $\beta\sim5.88$ for galaxy-sized haloes in the Millennium Run \cite{springel2005simulations}. Some analogous findings have been presented in the case of galaxy-sized haloes within the Aquarius simulations \cite{springel2008aquarius}.

Next, we consider the following function
\begin{align}\label{s6}
\gamma(\alpha,z)=\int_{0}^{z}\exp(-x)x^{\alpha-1}dx\quad (\textmd{R}
e~\alpha>0),
\end{align}
which is the lower incomplete Gamma function. This expression results from the splitting of the complete Gamma function \cite{prym1877theorie}
$\Gamma(\alpha)=\int_{0}^{\infty}\exp(-x)x^{\alpha-1}dx$.
Now, the total mass $M$ of the Einasto parameterization with density function
\eqref{s5}, can be written as
\begin{align}\label{s7}
M=4\pi(\rho_{0}+e_{0})\int_{0}^{\infty}x^{2}\exp\left(-u\right)dx.
\end{align}
where $u=\left(\frac{r}{h}\right)^{\frac{1}{\beta}}$. The above
expression can be rewritten as
\begin{align}\label{s8}
M=4\pi(\rho_{0}+e_{0})\beta\Gamma(3\beta), \quad \textmd{with} \quad
\Gamma(3\beta)=\int_{0}^{\infty}\exp(-u)u^{3\beta-1}du.
\end{align}
Furthermore, the value of the density profile in terms of the total charged
mass terms using Eq. \eqref{s5}, can be given as
\begin{align}\label{s9}
\rho(r)=\frac{M}{4\pi\beta
h^{3}\Gamma(3\beta)}\exp\left[-\left(\frac{r}{h}\right)^{\frac{1}{\beta}}\right].
\end{align}

\section{Matter Distribution and Einasto's Dark Matter Halos}

In this section, we describe the formation of a BH solution endowed with spherical symmetry
in the context of the Einasto density model. Furthermore, let $M$ be
the total mass of a gravitational object modeled by the density function \eqref{s9}, which reduces to the Gaussian distribution with $h=\sqrt{\theta}$ \textmd{and} $\beta=\frac{1}{2}$ studied in \cite{nicolini2006noncommutative} for the derivation of non-commutative geometry motivated by Schwarzschild BH. We consider the most generic form of static, spherically
symmetric metric as
\begin{align}\label{m1}
ds^{2}=g_{00}(r)dt^{2}-g_{00}^{-1}(r)dr^{2}-r^{2}d\Omega^{2},
\end{align}
with $d\Omega^{2}\equiv
d\theta^{2}+\sin^{2}\theta d\varphi^{2}$. The total stress-energy tensor is considered to be the sum of two
parts, $M^{\mu}_{\eta}$ and $E^{\mu}_{\eta}$, for matter and
electromagnetic interactions, respectively as
\begin{align}\label{m2}
T^{\mu}_{\eta}=M^{\mu}_{\eta}+E^{\mu}_{\eta}.
\end{align}
The usual expression of stress-energy tensor for anisotropic
matter distribution with density source \eqref{s9} can be written
in terms of a diagonal matrix as
\begin{align}\label{m3}
M^{\mu}_{\eta}=\left(
                                            \begin{array}{cccc}
                                              \rho & 0 & 0 & 0 \\
                                              0 & -p_{r} & 0 & 0 \\
                                              0 & 0 & -p_{\bot} & 0 \\
                                               0& 0 & 0 & -p_{\bot} \\
                                            \end{array}
                                          \right), \quad
                                            p_{r}\neq p_{\perp}.
\end{align}
Here, the functions $\rho$, $p_{r}$ and $p_{\bot}$ denote the energy
density, radial pressure, and tangential pressure, respectively. The
electromagnetic contributions can be expressed by the following
expression
\begin{align}\label{m4}
 E_{\eta}^{\mu}=-\frac{1}{4\pi}\left(F_{\eta\nu}F^{\mu\nu}-\frac{1}{4}\delta^{\mu}_{\eta}F_{\nu\sigma}F^{\nu\sigma}\right),
 \end{align}
 where $F_{\mu\eta}$ is the electromagnetic field tensor, which can be defined in terms of four potential
 $\mathcal{A}_{\mu}$
 as
\begin{align}\label{m5}
 F_{\mu\eta}=\frac{\partial \mathcal{A}_{\eta}}{\partial x^{\mu}}-\frac{\partial \mathcal{A}_{\mu}}{\partial x^{\eta}}.
 \end{align}
 In the rest frame of reference, we adopt the gauge field $\mathcal{A}_{\mu}$ as
 \begin{align}\label{m6}
 \mathcal{A}_{\mu}=(\phi(r),0,0,0).
 \end{align}
To determine the unknown metric coefficients appearing in
Eq. \eqref{m1}, we consider the Einstein-Maxwell field
equations, given as
\begin{align}\label{m7}
&R^{\mu}_{\eta}-\frac{1}{2}\mathbf{R}\delta^{\mu}_{\eta}=-8\pi
T^{\mu}_{\eta},
\\\label{m8}
&F_{\mu\eta;\nu}+F_{\eta\nu;\mu}+F_{\nu\mu;\eta}=0,
\\\label{m9}
&F^{\mu\eta}_{~~;\eta}=-4\pi J^{\mu}.
\end{align}
where $R_{\eta}^{\mu}$ is the Ricci tensor, $\mathbf{R}$ is the
Ricci scalar. {It is important to note that we use
the convention where fundamental constants like $G$ (gravitational
constant) and $c$ (speed of light) are set equal to $1$ (i.e.,
$G=c=1$)}. Further, the current density $J^{\mu}$ is can be defined
in terms of electric charge density $\sigma_{e}(r)$ as
\begin{align}\label{m10}
 J^{\mu}=\sigma_{e}(r)V^{\mu},
 \end{align}
where the four-velocity $V^{\mu}$ of the fluid is given as
\begin{align}\label{m11}
 V^{\mu}=\left(\frac{1}{\sqrt{\textsl{g}_{\mu\mu}}},0,0,0\right).
 \end{align}
 Then, combining Eq. \eqref{m1} with Eqs. \eqref{m9}-\eqref{m11} provide the following
 differential equation
\begin{align}\label{m12}
 \frac{d^{2}\phi}{dr^{2}}+\frac{2}{r}\frac{d\phi}{dr}=\frac{4\pi\sigma_{e}}{\sqrt{\textsl{g}_{00}}},
 \end{align}
 which is linear in $\phi$. Solving the above differential equation, we get
\begin{align}\label{m13}
 \frac{d\phi}{dr}=\frac{q(r)}{r^{2}\sqrt{\textsl{g}_{00}}},
 \end{align}
 where the total charge $q(r)$ is given as
\begin{align}\label{m14}
 \frac{d\phi}{dr}=\frac{q(r)}{r^{2}\sqrt{\textsl{g}_{00}}},\quad
 \textmd{with}\quad q(r)=4\pi\int^{r}_{0}\sigma_{e}(x)x^{2}dx.
 \end{align}
 Here, $q(r)$ is the total charge enclosed by the spherically
 symmetric gravitational source. Thus the non-null
 constituents of $M^{\mu}_{\eta}$ are the four diagonal elements
\begin{align}\label{m15a}
 E^{0}_{0}=E^{1}_{1}=-E^{2}_{2}=-E^{3}_{3}=\frac{q^{2}(r)}{8\pi r^{4}}.
 \end{align}
 Finally, the total stress-energy tensor for anisotropic charged matter
distribution is given as
\begin{align}\label{m16a}
T^{\mu}_{\eta}=\left(
                                            \begin{array}{cccc}
                                              \rho+\frac{q^{2}}{8\pi r^{4}} & 0 & 0 & 0 \\
                                              0 & -p_{r}+\frac{q^{2}}{8\pi r^{4}} & 0 & 0 \\
                                              0 & 0 & -p_{\bot}-\frac{q^{2}}{8\pi r^{4}} & 0 \\
                                               0& 0 & 0 & -p_{\bot}-\frac{q^{2}}{8\pi r^{4}} \\
                                            \end{array}
                                          \right),
\end{align}
where the total charge $q(r)$ and the electric field $\tilde{E}(r)$
are related by the formula
\begin{align}\label{m15}
\frac{q^{2}(r)}{8\pi r^{4}}=\frac{\tilde{E}^{2}(r)}{8\pi}.
 \end{align}
The Einstein-Maxwell field equations along with the conservation equation $\nabla _{\mu}T^{\mu\eta}=0$, produces the anisotropic hydrostatic equilibrium equation (the Tolman-Oppenheimer-Volkov equation) for the electrostatic case as
\begin{align}\label{m16}
 \frac{dp_{r}}{dr}=-\frac{1}{2\textsl{g}_{00}}\frac{d\textsl{g}_{00}}{dr}(\rho+p_{r})
 +\frac{2}{r}\left[\frac{q^{2}}{8\pi
 r^{3}}\frac{dq}{dr}-(p_{r}-p_{\bot})\right],
 \end{align}
 where
 \begin{align}\label{m17}
\frac{1}{2\textsl{g}_{00}}\frac{d\textsl{g}_{00}}{dr}=\frac{4\pi
r^{4}p_{r}+mr-q^{2}}{r(r^{2}-2mr+q^{2})},
 \end{align}
 and the mass function $m(r)$ is
 \begin{align}\label{m18}
m=4\pi\int^{r}_{0}\rho(x)x^{2}dx+\int^{r}_{0}\frac{q(x)}{x}\frac{dq(x)}{dx}dx.
 \end{align}
 We can retrieve the standard form of
 Tolman-Oppenheimer-Volkoff equation by setting $q=0$ in
 Eq.\eqref{m16}.
 In terms of the lower
 incomplete Gamma function, the gravitational mass $m(r)$ takes the form
 \begin{align}\label{mv18}
m=\frac{M}{\Gamma(3\beta)}\gamma\left(3\beta,\left(\frac{r}{h}\right)^{\frac{1}{\beta}}\right)
+\int^{r}_{0}\frac{q(x)}{x}\frac{dq(x)}{dx}dx.
 \end{align}
 We require that $T^{0}_{0}=T^{1}_{1}=-\rho(r)$. For this requirement, the source appears to be
 self-gravitational droplet with charged anisotropic fluid,
 whose radial and tangential pressures, respectively, are defined as
\begin{align}\label{m19}
p_{r}(r)-\frac{q^{2}}{8\pi r^{4}}=-\left(\rho(r)+\frac{q^{2}}{8\pi
r^{4}}\right)=\frac{M}{4\pi\beta
h^{3}\Gamma(3\beta)}\exp\left[-\left(\frac{r}{h}\right)^{\frac{1}{\beta}}\right].
 \end{align}
 \begin{align}\label{m20}
p_{\bot}+\frac{q^{2}}{8\pi r^{4}}=-\rho(r)\left[1-\frac{1}{2\beta}\left(\frac{r}{h}\right)^{\frac{1}{\beta}}\right]-\frac{q^{2}}{8\pi
r^{3}}\frac{dq}{dr}.
 \end{align}
 \begin{figure}
\centering{{\includegraphics[height=3 in, width=3 in]{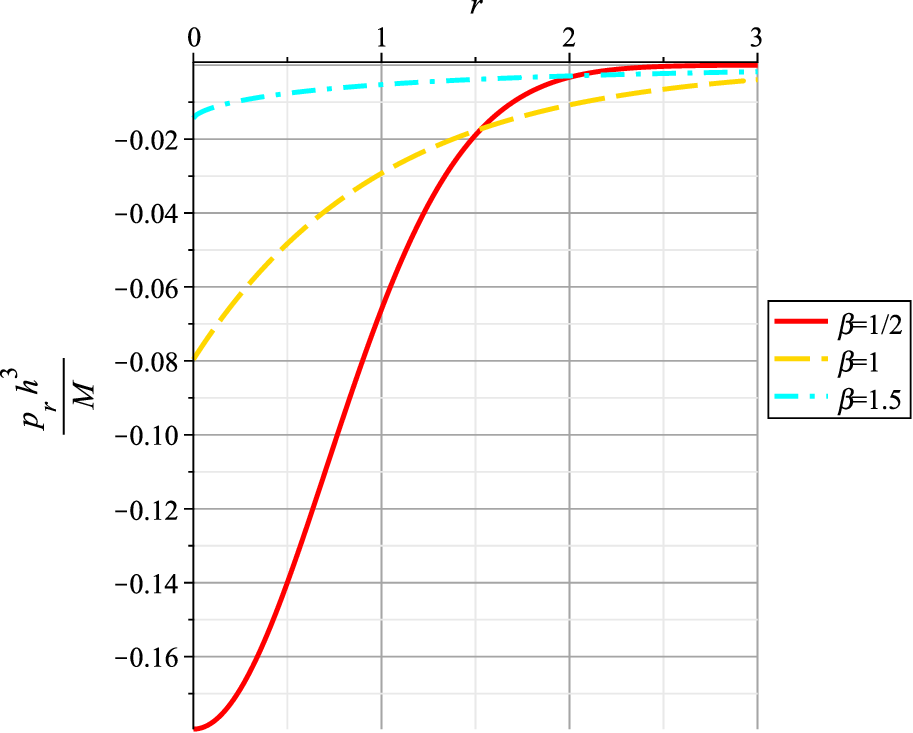}}}
\qquad{{\includegraphics[height=3 in, width=3 in]{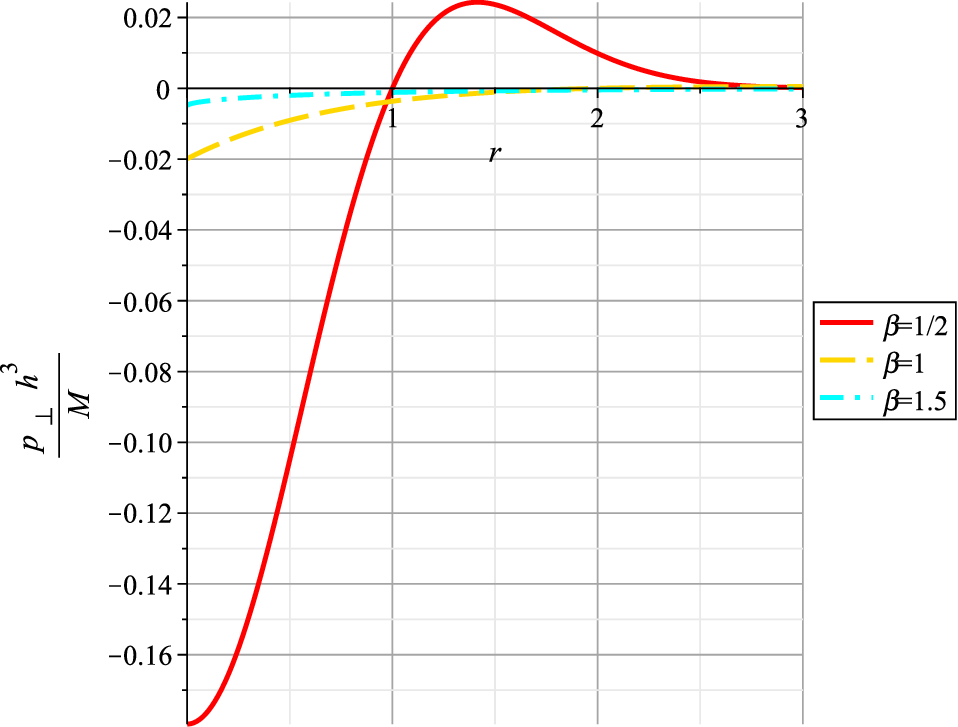}}}
\caption{{Behaviors of $p_{r}h^{3}/M$ (on the left panel) and
$p_{\bot}h^{3}/M$ (on the right panel) versus $r/h$ with de
Sitter-like EoS, $\rho=-p_{r}$}. In case of DM, Einasto mass
$M=4.57\times10^{9}M_{\odot}$, $h=2.121\times10^{-9}$ kpc and the
Einasto index $\beta=0.7072$
\cite{einasto1969galactic,einasto1969andromeda}. For normalized
units $|p_{r}(0)|=|p_{\perp}(0)|\approx5.8\times10^{-41}m^{-2}$ and
in S.I units
$|p_{r}(0)|=|p_{\perp}(0)|\approx7\times10^{3}\textmd{N/m}^{2}$. It
is significant to understand that the tangential pressure alters its
sign at $r_{0}\approx0.3$ kpc located within the active galactic
nucleus \cite{urry1995unified}.}\label{1s}
\end{figure}
The behavior of the radial (left panel) and tangential pressure
(right panel) components is described in FIG \ref{1s}. The most notable feature is the negative values of both the pressure components throughout the graph. This is a feature of exotic matter,
which is commonly associated with gravitationally bound stellar
objects. It is observed that $p_{r}$ and $p_{\bot}$ exhibit maximum
values near the horizon, while their values decrease as we move away
from it. At large radial distances, the graphs converge asymptotically to $0$, implying that the object's pressure gradually vanishes far from its core. By following this process, one can infer that the energy-momentum conservation equation is satisfied identically. Eventually, it is straightforward to show that Eq. \eqref{m16} is trivially satisfied for the values of $p_{r}$ and $p_{\bot}$ as provided by Eqs. \eqref{m19} and \eqref{m20}, respectively. This technique implies that the Einasto density model characterizes an anisotropic self-gravitational fluid. Now by solving the Einstein-Maxwell field equations, with density profile \eqref{m1}, matter source\eqref{m16}, and the values of $p_{r}$ and $p_{\bot}$, the line element turns out to be
\begin{align}\label{m21}
ds^{2}=\left(1-\frac{2m}{r}+\frac{q^{2}}{r^{2}}\right)dt^{2}-\left(1-\frac{2m}{r}+\frac{q^{2}}{r^{2}}\right)^{-1}dr^{2}
-r^{2}d\Omega^{2},
 \end{align}
or using expression \eqref{mv18}, we get
 \begin{align}\nonumber
ds^{2}&=\left(1-\frac{2M}{r\Gamma(3\beta)}\gamma\left(3\beta,\left(\frac{r}{h}\right)^{\frac{1}{\beta}}\right)
+\frac{q^{2}}{r^{2}}+\int^{r}_{0}\frac{q(x)}{x}\frac{dq(x)}{dx}dx\right)dt^{2}-\left(1-\frac{2M}{r\Gamma(3\beta)}\gamma\left(3\beta,
\left(\frac{r}{h}\right)^{\frac{1}{\beta}}\right)+\frac{q^{2}}{r^{2}}\right.
\\\label{m22}
&\left.+\int^{r}_{0}\frac{q(x)}{x}\frac{dq(x)}{dx}dx\right)^{-1}dr^{2}
-r^{2}d\Omega^{2},
 \end{align}
where
\begin{align}\label{m23}
\textsl{g}_{00}(r)=1+\left(\frac{q}{r}\right)^{2}-\frac{2M}{r\Gamma(3\beta)}\gamma\left(3\beta,
\left(\frac{r}{h}\right)^{\frac{1}{\beta}}\right)+\int^{r}_{0}\frac{q(x)}{x}\frac{dq(x)}{dx}dx.
 \end{align}
Here, it is significant to note that the metric \eqref{m22} reduces
to the classical Reissner-Nordstr\"{o}m line element in the limit
$\frac{r}{h}\rightarrow\infty$. Moreover, by assuming the metric
coefficient $g_{00}$ as a function of $ \frac{r}{h}$ and introducing
the rescaled mass $\omega:=\frac{M}{h}$, there must exists a value
of $\omega$, say, $\omega_{0}$, such that  $g_{00}$ has a double
root at $u_{0}:=\frac{r_{0}}{h}$. Furthermore, two distinct event
horizons exist for $\omega>\omega_{0}$ and no horizon for
$0<\omega<\omega_{0}$. Additionally, the behavior of the
relativistic metric potential $\textsl{g}_{00}$ may be realized by
using the relation (see \cite{abramowitz2006handbook} for details)
\begin{align}\label{m24}
\gamma(a,z)=z^{-a}\gamma^{\ast}(a,z)\Gamma(a,z), \quad \textmd{with}
\quad
\gamma^{\ast}(a,z)=\exp(-z)\sum^{\infty}_{n=0}\frac{z^{n}}{\Gamma(a+n+1)}.
 \end{align}
 The metric function $\textsl{g}_{00}(r)$ turns out to be
 \begin{align}\label{m25}
\textsl{g}_{00}(r)=1+\left(\frac{q}{r}\right)^{2}-
2\omega\left(\frac{r}{h}\right)^{2}\exp\left[-\left(\frac{r}{h}\right)^{\frac{1}{\beta}}\right]
\sum^{\infty}_{n=0}\frac{\left(\frac{r}{h}\right)^{\frac{n}{\beta}}}{\Gamma(3\beta+n+1)}
+\int^{r}_{0}\frac{q(x)}{x}\frac{dq(x)}{dx}.
 \end{align}
The above-mentioned expression behaves alternatively, as in the case
of Reissner-Nordstr\"{o}m BH and the Schwarzschild BH, where
singularity appears at $r=0$. In this case, a standard de Sitter
core is used to characterize the central region. Consequently, one
can deduce that Einasto's density model with anisotropic
energy-momentum tensor for an EoS $p_{r}=-\rho$, solves the problem
of central density. Furthermore, there is no naked singularity as
well as no event horizon for the case $\omega<\omega_{0}$.
\begin{figure}
\centering{{\includegraphics[height=2.5 in, width=5 in]{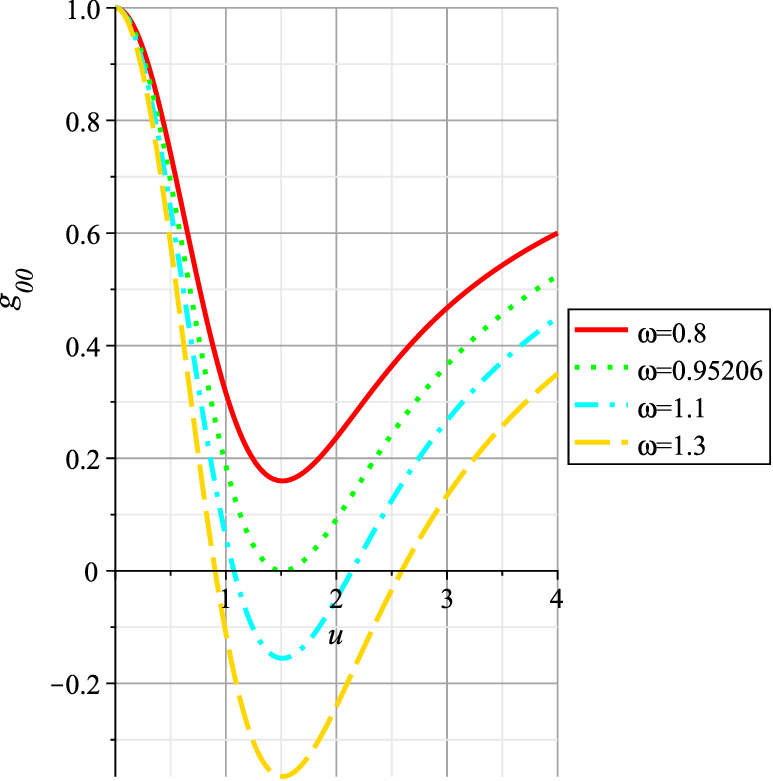}}}
\caption{Behavior of the relativistic metric potential
$\textsl{g}_{00}$ versus $u:=(\frac{r}{h})^{1/\beta}$ with Einasto
index $\beta=0.5$ for different values of the rescaled parameter
$\omega$.}\label{2s}
\end{figure}
FIG. \ref{2s} shows the variation of the metric function subject to
the Einasto density model. The metric function assumes a negative
values, a typical feature of a gravitational field. This shows that
the object has a gravitational attraction on the nearby test
particles. The radii of the event horizons in terms of the variable
$u$ are characterized by the intersections on the $u$-axis. There is
only one degenerate horizon for $\omega_{0}=2.28378$ with
$\omega=\omega_{0}=0.95206$. The value $\omega_=0.8<\omega_{0}$
corresponds to no horizon case, whereas $\omega=1.1> \omega_{0}$
corresponds to two horizons. Notable, the former case represents the
self-gravitational droplet coupled with anisotropic stress energy
tensor. Let us now calculate the Hawking temperature
$T_{\mathcal{H}}$ for this novel class of BHs by using the formula
\cite{nicolini2006noncommutative} as
 \begin{align}\label{m26}
T_{\mathcal{H}}=\frac{1}{4\pi}\left(\frac{d\textsl{g}_{00}}{dr}\right)_{r=r_{\mathcal{H}}}=\frac{2q}{r_{\mathcal{H}}^{2}}\left(\frac{dq}{dr}-\frac{q}{r_{\mathcal{H}}}\right)
+\frac{1}{4\pi
r_{\mathcal{H}}}\left\{1-\frac{r_{\mathcal{H}}^{3}\exp\left[-\left(\frac{r}{h}\right)^{\frac{1}{\beta}}\right]}{h^{3}\beta\gamma(3\beta,\left(\frac{r}{h}\right)^{\frac{1}{\beta}})}\right\}
+\frac{q}{r_{\mathcal{H}}}\frac{dq}{dr_{\mathcal{H}}},
 \end{align}
 where $M$ and $r_{\mathcal{H}}$ are the total mass and the position of the event horizon, respectively. We can obtain the eventual event horizon by using the expression $\textsl{g}_{00}(r_{\mathcal{H}})=0$, that is
 \begin{align}\label{ms27}
r_{\mathcal{H}}=2m-\frac{q^{2}}{r}=\frac{2M}{r\Gamma(3\beta)}\gamma\left(3\beta,\left(\frac{r}{h}\right)^{\frac{1}{\beta}}\right)-\frac{q^{2}}{r}-\frac{1}{r}\int^{r}_{0}\frac{q(x)}{x}\frac{dq(x)}{dx}.
\end{align}
It is notable that the expression \eqref{m26} reduces to the classical result
$T_{\mathcal{H}}=(4\pi r_{\mathcal{H}})^{-1}$ for the case $\frac{r_{\mathcal{H}}}{h}\gg1$ and $q=0$.

\section{The Effective Potential}

This section aims to figure out whether the diffuse BH models
comprised of DM, presented in the previous section, can be
consistent with the model of central galactic BH with mass
$M_{BH}=4.1\times10^{-6}M_{\odot}$ and Schwarzschild radius
$R_{BH}=2G_{N}M_{BH}/c^2=17.4~R_{\odot}=3.92\times10^{-7}~\textmd{pc}$.
For this, we need to understand the relevant parameters, i.e., the
Einasto index $\beta$ and the rescaling factor $h$ in the considered
model. More specifically, this can be completed in two ways.
Firstly, by imposing the condition that the total mass $M$ appearing
in the geometry \eqref{m22} coincides with $M_{BH}$. Nextly, when
measured at the $r_{\textsl{min}}$ of the effective potential
associated with a massive particle, the mass $m$ offers an accurate
estimation for $M_{BH}$. Alternatively, we require that

\begin{align}\label{m27}
1-\frac{m(r_{min})}{M_{BH}}\leq10^{-2}.
\end{align}
This expression can be rewritten in an equivalent form by using Eq. \eqref{m18} as
\begin{align}\label{m28}
\Delta\gamma\equiv\frac{1}{\Gamma(3\beta)}\gamma\left(3\beta,\left(\frac{r_{min}}{h}\right)^{\frac{1}{\beta}}\right)-0.99\geq0,
\end{align}
It turns out that the aforementioned criterion not only guarantees the equivalence between our diffused self-gravitating structure and the Schwarzschild effective potential at their minimum but they both also coincide within a substantial range of it as well as in the far limit. The effective potential $U^{\textsl{eff}}$ for this investigation is calculated by using the geodesic equation of motion. Particularly, by employing the method as discussed in \cite{fliessbach2012allgemeine}, we can transform the radial equation into an energy-conservation equation given as
\begin{align}\label{m29}
\frac{dr}{d\tau}+U^{\textsl{eff}}(r)=constant,
\end{align}
where $\tau$ denotes the proper time, which depends on whether a
massless or heavy particle is assumed. Therefore, the effective
potential corresponding to the spherical metric characterized by the
ansatz \eqref{m1} is given as \cite{fliessbach2012allgemeine}
\begin{align}\label{m30}
U^{\textsl{eff}}=\frac{\ell}{2r^{2}}\left(1-\frac{2m}{r}+\frac{q^{2}}{r^{2}}\right)-\xi\frac{m}{r},
\quad \textmd{where} \quad {\displaystyle
\xi={\begin{cases}+1&{\text{if }}m_{p}\neq0\\-1&{\text{if
}}m_{p}=0\\\end{cases}}},
\end{align}
with $\ell$ and $m_{p}$ represent the total angular momentum per
unit mass and the mass of a test particle, respectively. We can
express the above relation using Eq. \eqref{s6} as
\begin{align}\label{m31}
U^{\textsl{eff}}=\frac{\ell^{2}}{2r^{2}}-\frac{M_{BH}}{\Gamma(3\beta)}\gamma\left(3\beta,\left(\frac{r}{h}\right)^{2}\right)
\left(\frac{\xi}{r}+\frac{\ell^{2}}{r^{2}}\right)+\frac{\ell^{2}q^{2}}{2r^{4}}.
\end{align}
Notably, the effective potential for the Schwarzschild case can be
written as
\begin{align}\label{m32}
U^{\textsl{eff}}_{\textsl{S}}=\frac{\ell^{2}}{2r^{2}}-{M_{BH}}\left(\frac{\ell^{2}}{r^{2}}+\frac{\xi}{r}\right).
\end{align}
Let us take $r_{s}=2M_{BH}$ and consider a rescaling of the
variables $\ell$ and $r$ such that $\mathscr{L}=\frac{\ell}{r_{s}}$
and $r^{\ast}=\frac{r}{r_{s}}$. Thus for the massive case, the Schwarzschild effective potential may take the following form
\begin{align}\label{m33a}
U^{\textsl{eff}}_{\textsl{S}}=-\frac{1}{2r^{\ast}}+\frac{\mathscr{L}^{2}}{2r^{\ast2}}\left(1-\frac{1}{r^{\ast}}\right),
\end{align}
where $r^{\ast}=1$ is defined as the location for the gravitational event horizon. In addition, the maximum and minimum values of $r^{\ast}$ are given as
\begin{align}\label{m34a}
r^{\ast}_{\textsl{min}}=\frac{\mathscr{L}^{2}}{2}\left(1+\sqrt{1-\frac{3}{\mathscr{L}^{2}}}\right),
\quad
r^{\ast}_{\textsl{max}}=\frac{\mathscr{L}^{2}}{2}\left(1+\sqrt{1-\frac{3}{\mathscr{L}^{2}}}\right) \quad \textmd{with } \quad \mathscr{L}>\sqrt{3},
\end{align}
respectively. After introducing the above-mentioned rescaling for
$\ell$ and $r$ in Eq. \eqref{m30}, we obtain
\begin{align}\label{m33}
U^{\textsl{eff}}=\frac{\mathscr{L}^{2}}{2r^{2}}-\frac{1}{\Gamma(3\beta)}
\gamma\left(3\beta,\left(\frac{r^{\ast}}{\mathcal{H}}\right)\right)\left(\frac{1}{2r^{\ast2}}
+\frac{\mathscr{L}^{2}}{2r^{\ast3}}\right) \quad \textmd{with} \quad
\mathcal{H}=\frac{h}{r_{s}},
\end{align}
and
\begin{align}\label{m34}
\Delta\gamma\equiv\frac{1}{\Gamma(3\beta)}\gamma\left(3\beta,\left(\frac{r_{\textsl{min}}^{\ast}}{\mathcal{H}}\right)\right)-0.99\geq0.
\end{align}
This constraint represents an inequality in the Einasto index
$\beta$ and the parameter $\mathcal{H}$. To display that
the solution set obtained from this is non-empty, we will analyze
certain values of the parameter $\mathcal{H}$ to ensure the similar
magnitude orders of the associated parameter $h=r_{s}\mathcal{H}$ as
in \cite{einasto1969galactic,einasto1969andromeda,de2019estimation}.
We can solve \eqref{m34} by considering different values of the
$\mathcal{H}$ with respect to the Einasto index $\beta$. The
above-mentioned approach implies that we need to fix the value of
total angular momentum $\mathscr{L}$ because the specific value of
$r_{\textsl{min}}^{\ast}$ depends on $\mathscr{L}$. For example, the
value of the rescaled parameter
$h_{E}=2.121\times10^{-9}~\textmd{kpc}$ in the case of DM haloes
\cite{einasto1969galactic,einasto1969andromeda}. Therefore, the
value of the rescaled parameter in our case is
$h=3.92\times10^{-9}~\textmd{kpc}$  corresponding to
$\mathcal{H}=10$. We explore several choices of $\mathscr{L}$ and
$r_{\textsl{min}}^{\ast}$ to determine which values will fulfill Eq.
\eqref{m34}. We notice that the expression \eqref{m34} cannot be
fulfilled with $r_{\textsl{min}}^{\ast}=6$ and $\mathscr{L}=2$
because in this case, numerical results show that  $\Delta\gamma<0$
within the range $10^{-6}\leq\beta\leq13$ \cite{batic2021fuzzy}.
To address the issue that whether the presented
solution characterizes a fuzzy droplet or the fuzzy BH model with
$\mathcal{H}=10$ and considered $\beta$ so that Eq. \eqref{m34} is
fulfilled, we observe that $r_{s}=2M_{BH}$ implies
$M_{BH}=\frac{r_{s}}{2}$ in case of geometric units and the
rescaling mass parameter $\omega$ for our model can be defined as
\begin{align}\label{m35}
\omega=\frac{1}{2\mathcal{H}}=\frac{M_{BH}}{h}=\frac{r_{s}}{2h}
\quad \textmd{where} \quad h=r_{s}\mathcal{H}.
\end{align}
\begin{figure}
\centering{{\includegraphics[height=3 in, width=3.8 in]{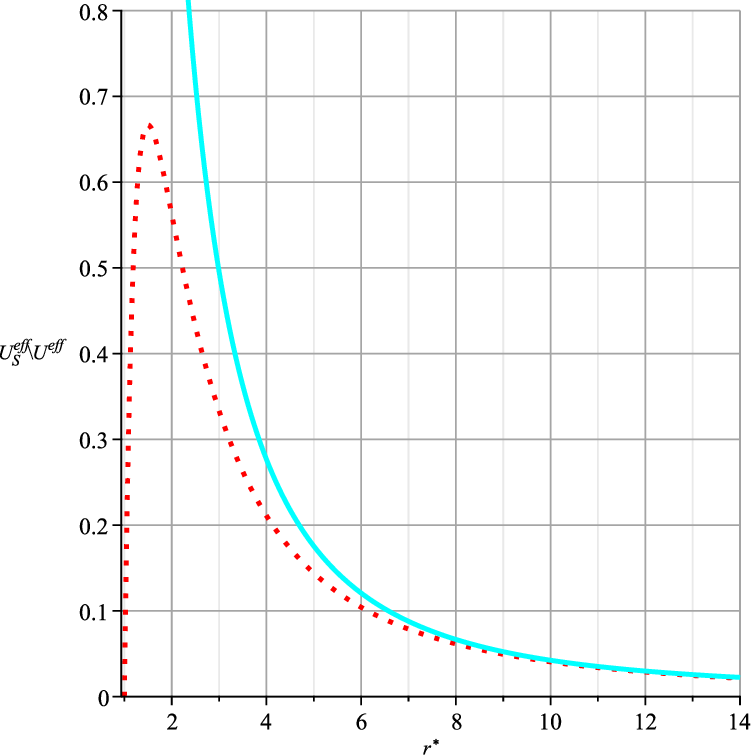}}}
\caption{Pictorial representation of the effective potentials
$U^{\textsl{eff}}_{S}$ (dotted line) and  $U^{\textsl{eff}}$ (solid)
with $\beta=0.2$, $\mathscr{L}=3$ and $\mathcal{H}=10$ for the
massless case. For this value of $\mathcal{H}$, the central galactic system is modeled as an anisotropic self-gravitational droplet with scaling factor $h$ of the order $h_{E}=2.21\times10^{-9}$kpc
\cite{einasto1969galactic,einasto1969andromeda}. Significantly, the
Schwarzschild BH's event horizon is located at $r^{\ast}$, whereas
the maximum value of $U^{\textsl{eff}}_{S}$ is positioned at the
photon sphere radius $r^{\ast}_{\gamma}=3/2$. The self-gravitational
droplet does not contain the photon sphere because
$U^{\textsl{eff}}$ fails to display a maximum value.}\label{3s}
\end{figure}
\begin{figure}
\centering{{\includegraphics[height=3 in, width=3.8 in]{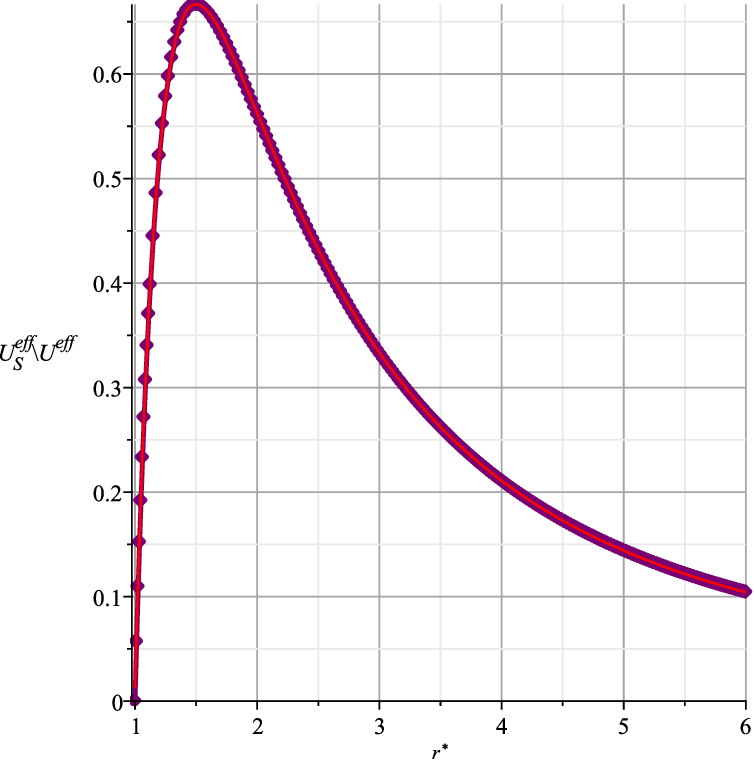}}}
\caption{Diagrammatic scheme of the effective potentials
$U^{\textsl{eff}}_{S}$ (point line) and  $U^{\textsl{eff}}$ (solid
line) with $\beta=1$, $\mathscr{L}=3$ and $\mathcal{H}=0.1$ for the
massless case. It is worth mentioning that for both the BH models
the event horizon is located at $r^{\ast}=1$ and the photon sphere
for both the potential ($U^{\textsl{eff}}_{S}$ and
$U^{\textsl{eff}}$) is positioned at
$r^{\ast}_{\gamma}=3/2$.}\label{4s}
\end{figure}
The behavior of effective potential for the Schwarzschild case and
the diffused compact droplet with different values of $\mathcal{H}$,
for the massless case, is illustrated in FIGs. \ref{3s} and
\ref{4s}. The profiles show that both the Schwarzschild effective
potential and the effective potential of our diffused
self-gravitational structure have the same photon sphere, and they
coincide over a significant range and at asymptotically large
distances. Furthermore, by introducing the rescaling
$r^{\ast}=\frac{r}{r_{s}}$ and $Q=\frac{q}{r_{s}}$ in expression
\eqref{m25}, the value of $\textsl{g}_{00}$ takes the form
\begin{align}\label{m36}
\textsl{g}_{00}=1+\left(\frac{Q}{r^{\ast}}\right)^{2}-
2\omega\left(\frac{r^{\ast}}{\mathcal{H}}\right)^{2}\exp\left[-\left(\frac{r^{\ast}}{\mathcal{H}}\right)^{\frac{1}{\beta}}\right]
\sum^{\infty}_{n=0}\frac{\left(\frac{r^{\ast}}{\mathcal{H}}\right)^{\frac{n}{\beta}}}{\Gamma(3\beta+n+1)}+\int^{r^{\ast}}_{0}\frac{Q(x)}{x}\frac{dQ(x)}{dx}.
 \end{align}
\begin{figure}
\centering{{\includegraphics[height=3 in, width=3.8 in]{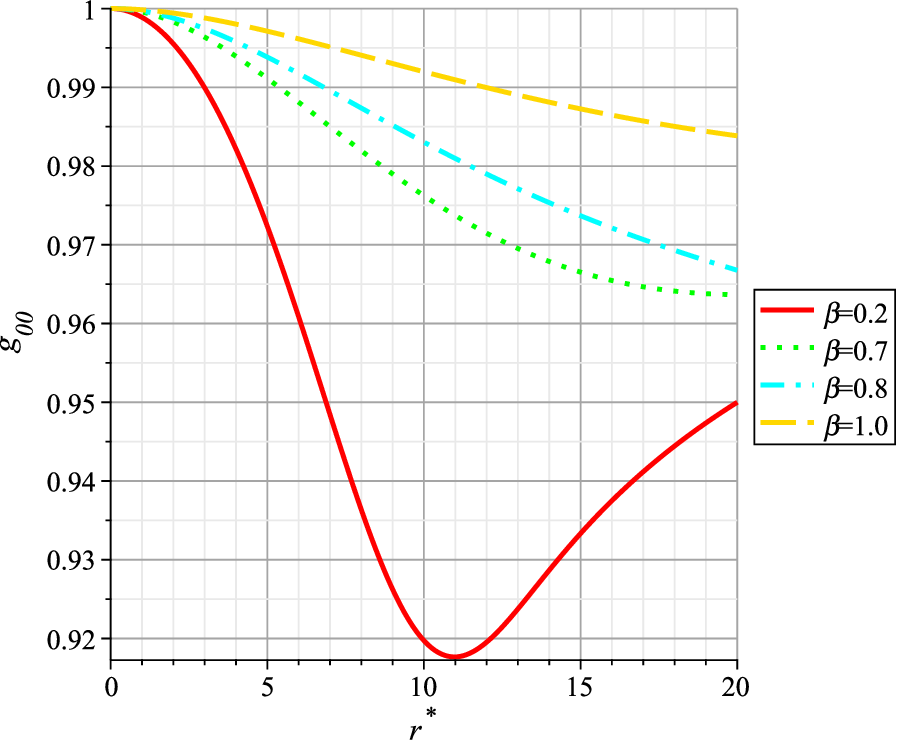}}}
\caption{Behavior of $\textsl{g}_{00}$ versus $r^{\ast}$ with
$\mathcal{H}=10$ for various values of $\beta$.}\label{5s}
\end{figure}
FIG. \ref{5s} reveals that the equation $\textsl{g}_{00}(r^{\ast})$
has no real roots. This shows that this model predicts a fuzzy
self-gravitating droplet.
\begin{figure}
\centering{{\includegraphics[height=3 in, width=3.8 in]{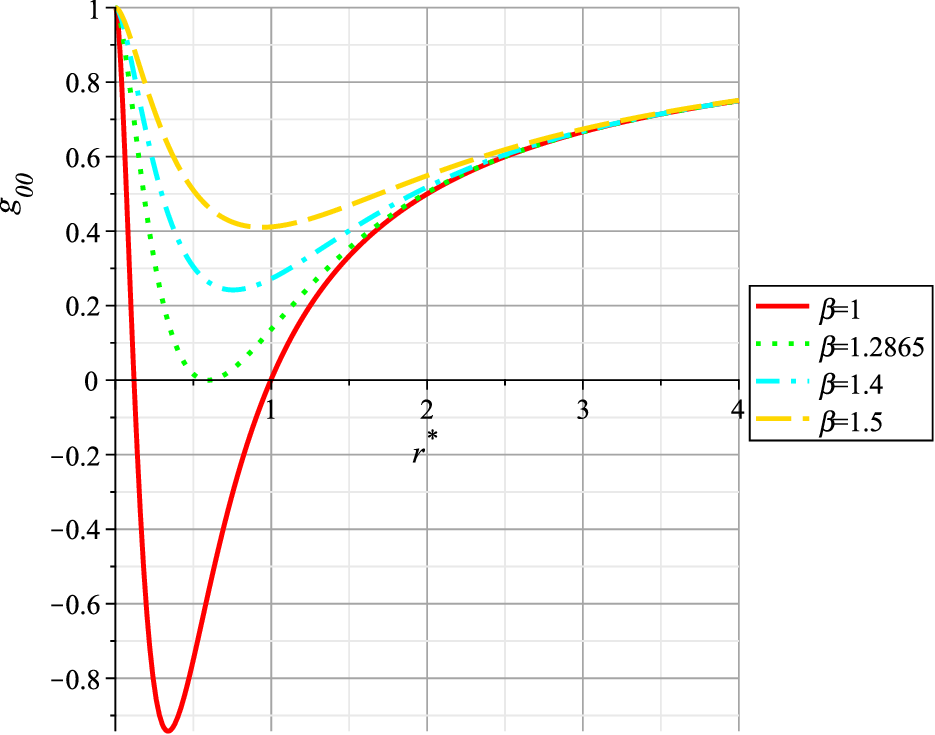}}}
\caption{Behavior of $\textsl{g}_{00}$ versus $r^{\ast}$ with
$\mathcal{H}=0.1$ for various values of $\beta$. In this case, the
event horizon is located at $r^{\ast}=1$, with $\beta=1$, and is
compatible with the standard Schwarzschild BH. However, the extreme
BH formed for the case $\beta=1.2865$ whose event horizon exists
at $r_{e}=2.33\times10^{-7}$pc.}\label{6s}
\end{figure}
A closer look at FIG. \ref{6s} suggests that, unlike in the case
when $\mathcal{H}= 10$, we have a more complex scenario. If
$\beta$ is less than $1.2865$, so we get a BH composed of DM with two
horizons. If $\beta$ is higher than $1.2865$, we observe a diffuse
DM droplet without a horizon. This same scenario occurs if we lower
the value of $\mathcal{H}$ even further.
\begin{figure}
\centering{{\includegraphics[height=3 in, width=3 in]{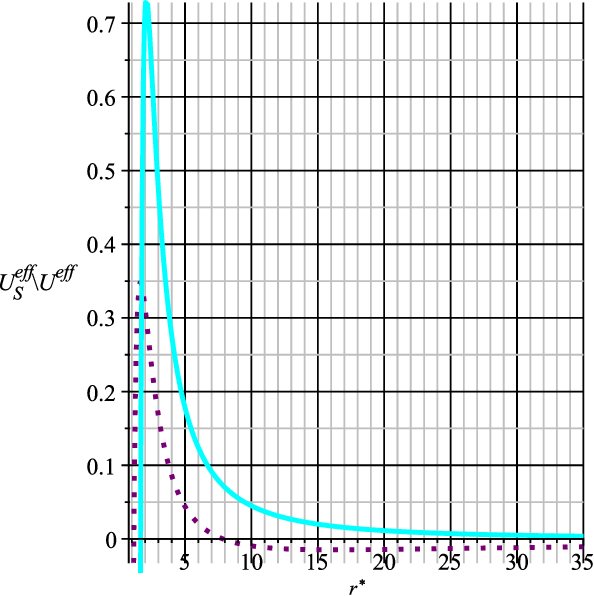}}}
\qquad{{\includegraphics[height=3 in, width=3 in]{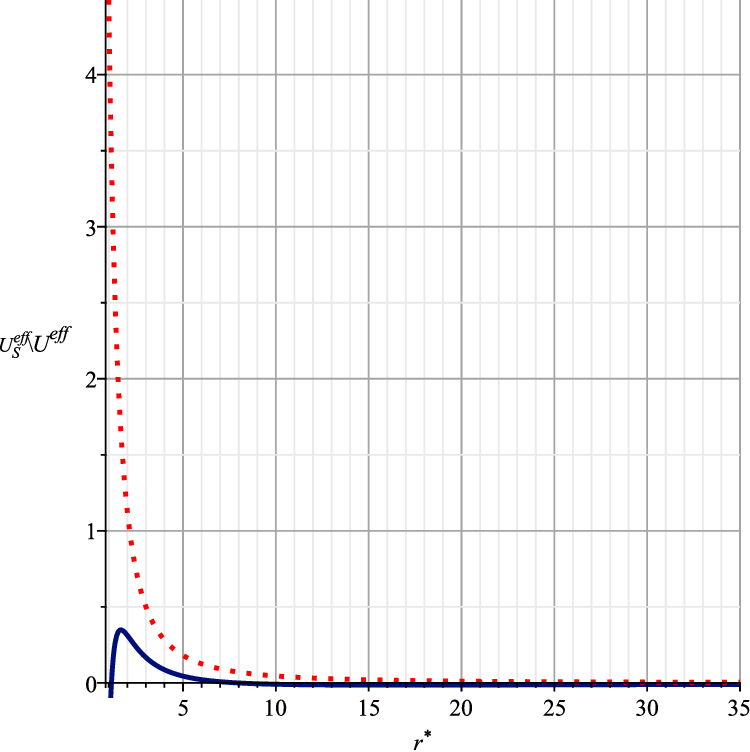}}}
\caption{Behavior of $U^{\textsl{eff}}_{S}$ (dotted line) and
$U^{\textsl{eff}}$ (solid line) versus $r^{\ast}$ on the left panel
with $\mathcal{H}=10$, $\mathscr{L}=3$ and the Einasto index
$\beta=0.2$ for the massive case. Plot of $U^{\textsl{eff}}_{S}$
(solid line) and $U^{\textsl{eff}}$ (dotted line) versus $r^{\ast}$
on the right panel with with $\mathcal{H}=0.1$, $\mathscr{L}=3$ and
the Einasto index $\beta=0.1$ for the massive case. The value of he
rescaled quantity $\mathcal{H}$ is selected to generate a scaling
factor $h$, comparable in magnitude to the one considered in
\cite{einasto1969galactic,einasto1969andromeda}. Also,
$U^{\textsl{eff}}_{S}(r^{\ast}_{\textsl{min}})=-0.01477$ and
$U^{\textsl{eff}}(r^{\ast}_{\textsl{min}})=-0.01477$, with
$r^{\ast}_{min}\approx16.35$. That is, both the potential
($U^{\textsl{eff}}_{S}$ and $U^{\textsl{eff}}$) have a common
minimum, and $U^{\textsl{eff}}$ provides a reliable approximation of
$U^{\textsl{eff}_{S}}$ within the neighborhood of
$r^{\ast}_{\textsl{min}}$.}\label{7s}
\end{figure}
The behavior of effective potential for the Schwarzschild case and
the diffused compact droplet with different values of $\mathcal{H}$
for the massive case is illustrated in FIGs. \ref{7s}. This shows
that the Schwarzschild effective potential and the effective potential of the proposed model have the same minimum. However, they
also agree in a broad neighborhood when $\mathcal{H} = 10$.

\section{Non-local Fuzzy Self-gravitational Diffused Dark Matter Models}

 By local self-gravitational models, we mean a usual EoS where the radial pressure and energy density are related at a specific point within the fluid distribution, i.e., $p_{r}(r)=-\rho(r)$. However, the gravitational models where stress-energy tensor constituents do not depend only on the metric, but also functionality extends by taking the average value of energy density throughout the remainder of the distribution, are known as non-local models.
Previously, we considered the possibility of developing local
self-gravitational anisotropic DM models using a de Sitter-type EoS
where the radial pressure has a particular form,
$\rho(r)=-p_{r}(r)$. Using the above-stated conditions, the Einasto
parameterization may lead to self-gravitational droplets or regular
BH distributions governed by the reformulated mass parameter
$\omega_{\textsl{m}}$. The supposition of non-local EoS with
anisotropic fluid configurations has been used in various studies
for comprehending the development of compact stellar objects through
the mechanism of general relativity
\cite{hernandez1999non,abreu2007sound}. Particularly, we consider an
Einasto profile-based charged anisotropic fluid characterized by a
non-local EoS whose static limit has the form
\cite{hernandez2004nonlocal}
\begin{align}\label{m37}
p_{r}(r)-\frac{q^{2}}{8\pi r^{4}}=\rho(r)+\frac{q^{2}}{8\pi
r^{4}}-\frac{2}{r^{3}}\int^{r}_{0}x^{2}\left(\rho(x)+\frac{q^{2}}{8\pi
r^{4}}\right)dx.
 \end{align}
 This equation signifies the local behavior of $\rho(r)$ and $p_{r}(r)$. The diffusive behavior of the Einasto density model appears to indicate that non-locality might be important by taking into account the fact that the changes in $\rho(r)$ imply the changes in the $p_{r}(r)$ through the whole volume. On substituting the density profile \eqref{s9}, the above relation become
 \begin{align}\label{m38}
p_{r}(r)-\frac{q^{2}}{8\pi r^{4}}=\frac{M}{4\pi\Gamma(3\beta)}\left\{\frac{\exp\left[-\left(\frac{r}{h}\right)^{\frac{1}{\beta}}\right]}{3\beta}-
\frac{2}{r^{3}}\gamma\left(3\beta,\left(\frac{r}{h}\right)^{\frac{1}{\beta}}\right)\right\}.
 \end{align}
 We are only concerned with the anisotropic matter distribution at hydrostatic equilibrium. Therefore, the fact that $p_{r}(r)>0$ within a region of finite size but  $p_{r}(r)<0$  outside allows us to present the effective size $R$ of the self-gravitating structure by imposing the constraint $p_{r}(R)=0$. Since $\rho\neq0$ within the region $R<r$, therefore the radius of this type of self-gravitational system will not be finite. We can find $R$ numerically by plotting $p_{r}(r)h^{2}$ against   $u=\left(\frac{r}{h}\right)^{\frac{1}{\beta}}$. In this respect, let us consider a rescaling of the mass parameter as $\omega=\frac{M}{h}$, thereby Eq. \eqref{m38} takes the following form
 \begin{align}\label{m39}
\left(p_{r}-\frac{q^{2}}{8\pi r^{4}}\right)\frac{h^{2}}{\omega}=\frac{M}{4\pi\Gamma(3\beta)}\left[\frac{\exp(-u)}{\beta}-\frac{2}{u^{3\beta}}\gamma(3\beta,u)\right],
 \end{align}
 which by using Eq. \eqref{m24} turns out to be
 \begin{align}\label{m40}
\left(p_{r}-\frac{q^{2}}{8\pi r^{4}}\right)\frac{h^{2}}{\omega}=\frac{\exp(-u)}{4\pi}\left[\frac{1}{\beta\Gamma(3\beta)}-2\sum^{\infty}_{n=0}\frac{u^{n}}{\Gamma(3\beta+n+1)}\right].
 \end{align}
 \begin{figure}
\centering{{\includegraphics[height=3 in, width=3.8 in]{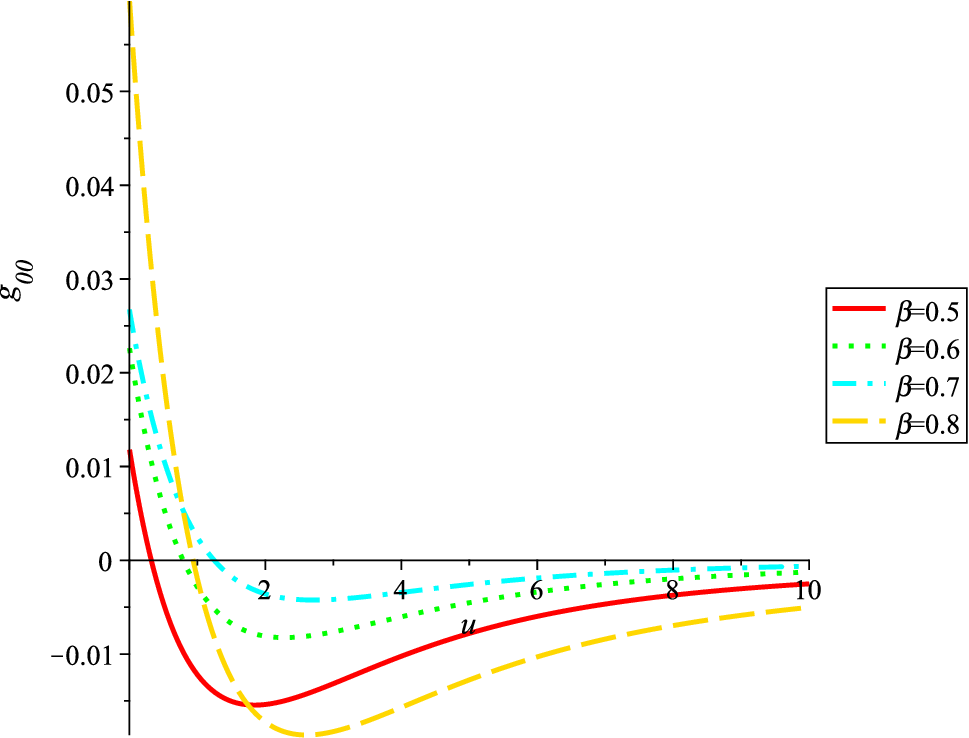}}}
\caption{Diagrammatic scheme of the rescaled form of
$\left(p_{r}-\frac{q^{2}}{8\pi r^{4}}\right)\frac{h^{2}}{\omega}$
against the variable $u$ for various values of $\beta$. In the
interior region $\left(p_{r}-\frac{q^{2}}{8\pi
r^{4}}\right)\frac{h^{2}}{\omega}>0$ and vanishes for certain value
of $u$ corresponding to the specific selection of parameter $\beta$.
However, $\left(p_{r}-\frac{q^{2}}{8\pi
r^{4}}\right)\frac{h^{2}}{\omega}<0$ and displays a minimum beyond
the value of $u$.}\label{8s}
\end{figure}
Furthermore, in FIG. \ref{8s}, we display the rescaled radial
pressure to demonstrate that it is positive within a given region
but negative beyond it. However, this does not imply that the
gravitational object has a fixed radius because the energy density
does not fall to zero outside of this region. In the next step, we examine a configuration of time-independent metric endowed with spherical symmetry coupled with a charged, anisotropic fluid \eqref{m16} subject to the non-local EoS \eqref{m37}. However, unlike the prior section, we adopt the following ansatz for the metric subject to the Einasto density profile
 \begin{align}\label{m41}
ds^{2}=A^{2}(r)dt^{2}-B(r)^{-1}dr^{2}-r^{2}d\Omega^{2}.
 \end{align}
Here, it is significant to highlight the fact that $B(r)$ coincides with $\textsl{g}_{rr}(r)=\textsl{g}_{00}^{-1}(r)$ emerging from the metric \eqref{m1}. This immediately suggests that the metric function $B(r)$ will share a similar investigation of the roots of $\textsl{g}_{rr}(r)$ in Sec. as discussed in \textbf{III}. Then, the Einstein-Maxwell field equations \eqref{m7} for this
 metric yield the following non-zero components
\begin{align}\label{m42}
\frac{B'}{r}-\frac{1-B}{r^{2}}=-8\pi\left(\rho+\frac{q^{2}}{8\pi
r^{4}}\right),
 \end{align}
 \begin{align}\label{m43}
\frac{2}{r}\frac{A'B}{A}-\frac{1-B}{r^{2}}=8\pi\left(p_{r}-\frac{q^{2}}{8\pi
r^{4}}\right),
 \end{align}
 \begin{align}\label{m44}
A'\frac{B}{A}+A''\frac{B}{A}+\frac{B'}{2r}+\frac{A'B'}{2A}=8\pi\left(p_{\bot}+\frac{q^{2}}{8\pi
r^{4}}\right),
 \end{align}
 where primes encode $r$-derivative. The hydrostatic equilibrium equation for the
 considered metric \eqref{m41} is given by
\begin{align}\label{m45}
p_{r}'=-(\rho+p_{r})\frac{A'}{A}-\frac{2}{r}\left[\frac{q^{2}}{8\pi
r^{3}}q'-(p_{r}-p_{\bot})\right].
 \end{align}
Finally, the integration of Eqs. \eqref{m42} and \eqref{m43}, respectively provides
\begin{align}\label{m46}
B(r)=1-\frac{2m}{r^{2}}+\frac{q^{2}}{r^{2}},
 \end{align}
\begin{align}\label{m47}
A^{2}(r)=\exp(\Phi(r)), \quad \Phi(r)=\int\Psi(r)dr \quad
\textmd{with} \quad
\Psi(r)=\frac{1}{B}\left[\frac{1}{r}\left(\frac{2m}{r}+\frac{q^{2}}{r^{2}}\right)+8\pi
r\left(p_{r}+\frac{q^{2}}{8\pi r^{4}}\right)\right].
 \end{align}
 In the above relation, the value of the mass function $m$ is provided by \eqref{mv18}, whereas the value of $p_{\bot}$ may be derived from \eqref{m45} with the help of \eqref{m47}.
 \begin{align}\label{m48}
p_{\bot}=p_{r}+\frac{r}{2}\left\{p_{r}'+\frac{\rho+p_{r}}{B}
\left[\frac{1}{r}\left(\frac{m}{r}-\frac{q^{2}}{2r^{2}}\right)+4\pi
r\left(p_{r}+\frac{q^{2}}{8\pi
r^{4}}\right)\right]+\frac{2}{r}\left(\frac{q'q^{2}}{8\pi
r^{3}}\right)\right\}.
 \end{align}
This expression shows that $B(r)$ is present in the denominator,
which shows that $p_{\bot}$ will turns singular at the roots of
$B(r)$. Finally, by using the relation \eqref{m47}, the metric
\eqref{m41} can be transformed in terms of two independent functions
$\exp(\Phi(r))$ and $m(r)$ as \cite{batic2021fuzzy}
 \begin{align}\label{m49}
ds^{2}=\exp(\Phi(r))dt^{2}-\left(1-\frac{2m}{r}+\frac{q^{2}}{r^{2}}\right)^{-1}dr^{2}-r^{2}d\Omega^{2}.
 \end{align}
 Here, the functions $\exp(\Phi(r))$ and $m(r)$ are classified as redshift function and shape function, as discussed in \cite{morris1988wormholes,visser1992dirty}. With an anisotropic stress-energy tensor, the above metric fulfills the following wormhole conditions \cite{nicolini2009noncommutative}.
 \begin{itemize}
   \item The value of tangential pressure $p_{\bot}(r)$ remains finite, i.e., $p_{\bot}(r)<\infty$.
   \item The metric coefficient $B(r)$ has two roots.
 \end{itemize}
 The investigation regarding the viability of this type of wormhole can be considered in future projects. We will avoid the later condition by imposing $\omega<\omega_{0}$, which ensures that the function $\Phi(r)$ is regular everywhere. This condition limits the metric variable $B(r)$ from entering in Eq. \eqref{m47} and possessing the real zeroes. As a result, we can deduce that the metric \eqref{m49} outlines a non-local self-gravitational fuzzy DM droplet.
 By introducing the variable $u=\left(\frac{r}{h}\right)^{\frac{1}{\beta}}$,
 the expression \eqref{m48} turns out to be
\begin{align}\label{m50}
\frac{h^{2}}{\omega}\left(p_{\bot}-\frac{q^{2}}{8\pi
r^{4}}\right)=\frac{1}{4\pi\Gamma(3\beta)}\left[\frac{\gamma(3\beta,u)}{u^{3\beta}}-\frac{u\exp(-u)}{2\beta^{2}}\right]
+\frac{\omega u^{2\beta}}{\Gamma^{2}(3\beta)\left(1-\frac{2\omega\gamma(3\beta,u)}{u^{\beta}\Gamma(3\beta)}\right)
}\left[\frac{\gamma(3\beta,u)}{u^{3\beta}}-\frac{\exp(-u)}{\beta}\right]^{2}.
 \end{align}
\begin{figure}[H]
\centering{{\includegraphics[height=3 in, width=3.8 in]{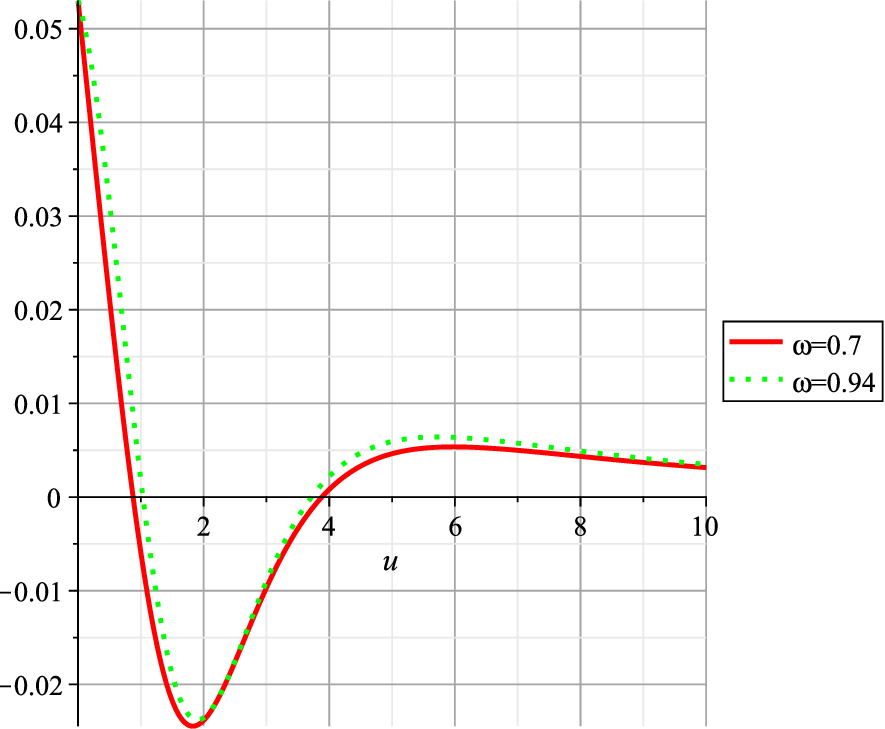}}}
\caption{Pictorial representation of the rescaled pressure
$\frac{h^{2}}{\omega}\left(p_{\bot}-\frac{q^{2}}{8\pi r^{4}}\right)$
with Einasto index $\beta=1/2$ for various values of the parameter
$\omega<\omega_{0}=0.5206$. The above representation has been
attained by making use of the expansion formula for the lower
incomplete Gamma function given in \eqref{m24}.}\label{9s}
\end{figure}
{The behavior of rescaled tangential pressure for two different values of the parameter $\omega$ is described in FIG. \ref{9s}. Both curves exhibit negative tangential pressure suggesting the presence of exotic matter, such as that found in self-gravitational compact systems. The tangential pressure
diminishes as the distance between curves rises. The parameter
$\omega$ affects the pressure distribution, with greater values
resulting in a flatter profile.}
\begin{figure}
\centering{{\includegraphics[height=3 in, width=3.8 in]{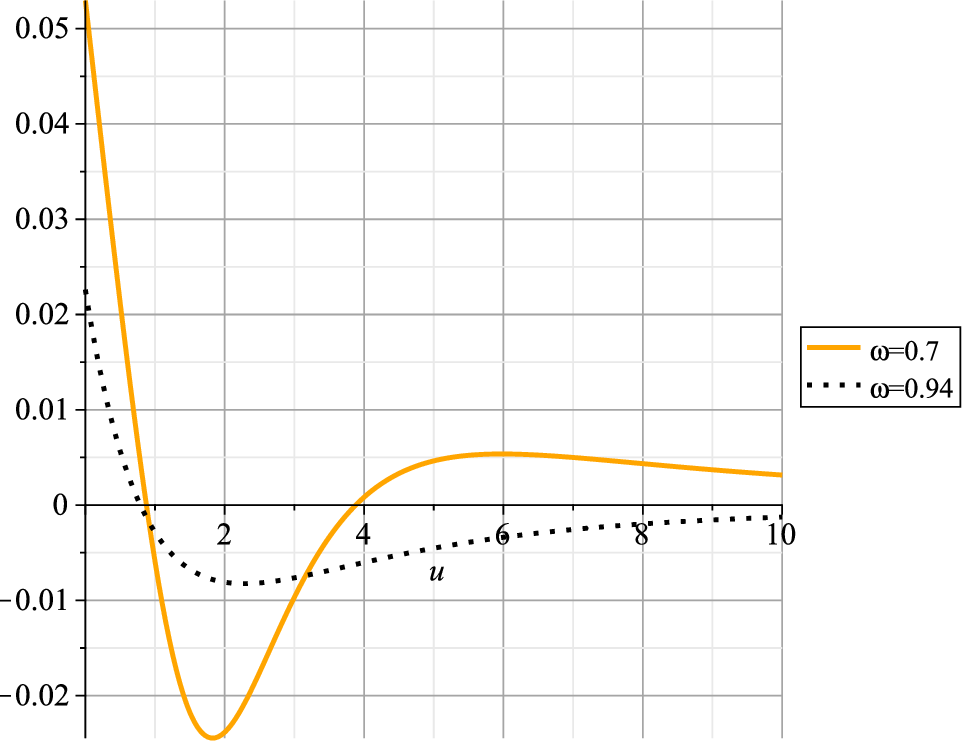}}}
\caption{Pictorial representation of
$\frac{h^{2}}{\omega}\left(p_{\bot}-\frac{q^{2}}{8\pi r^{4}}\right)$
for $\omega=0.94$ (solid line) and
$\frac{h^{2}}{\omega}\left(p_{r}-\frac{q^{2}}{8\pi r^{4}}\right)$
(dotted line) for $\omega=0.7$, with $\beta=1/2$ for different
values of the parameter $\omega<\omega_{0}=0.5206$. The above
representation has been attained by making use of the expansion
formula for the lower incomplete Gamma function given in
\eqref{m24}.}\label{10s}
\end{figure}
In FIG. \ref{10s}, we plot the rescaled radial and tangential
pressures for the same value of $\beta$ and rescaled mass. It is
observed that even for a nonlocal EoS, the self-gravitating droplet
does not exhibit any singularity at $r=0$. This fact can be easily
confirmed by evaluating the Kretschmann scalar $\mathcal{K}$ for the
metric \eqref{m49} as
\begin{align}\label{m51}
\mathcal{K}=R^{\mu\eta\nu\varsigma}R_{\mu\eta\nu\varsigma}=\frac{2}{r^{2}}\left(\Psi^{2}B^{2}+B'^{2}\right)+\frac{1}{4}\left(
\Psi^{2}B+2\Psi' B+\Psi B'\right)^{2}.
 \end{align}
The presence of $r^{-2}$  term in this relation does not provide any
guarantee that the Kretschmann scalar is singularity-free at $r=0$
for the assumed Einasto density model. However, by using the
expression \eqref{m24} and taking the limit $r\rightarrow0$, we get
 \begin{align}\label{m52}
\lim_{r\rightarrow0}R^{\mu\eta\nu\varsigma}R_{\mu\eta\nu\varsigma}=\frac{1}{9h^{6}\beta^{4}\Gamma^{4}(3\beta)}[
32M^{2}\beta^{2}\Gamma^{2}(3\beta)+\left(2\beta^{3}+\beta^{2}+\Gamma^{2}(3\beta)
+2\beta\Gamma^{2}(3\beta)\right)].
\end{align}
Thus the above expression is singularity-free at $r=0$. It can be easily shown that at space-like infinity, the metric \eqref{m49} displays the asymptotic nature at $B(r)\rightarrow 1$, whereas the asymptotic nature of the function $\Phi(r)$ can be obtained by using \eqref{m24} as
\begin{align}\label{m53}
\Phi(r)=\frac{2\beta M}{(1+\beta)r}+\cdot\cdot\cdot~,
\end{align}
where we have neglected the factors that are exponentially
decreasing. This implies that the metric \eqref{m49} is
particularized by asymptotically flat manifold because
$\exp(\Phi(r))\rightarrow1$, as $r\rightarrow\infty$. This section
can be concluded by describing that the compact gravitational droplet
enables the bound states of massive particles. Moreover, the
effective potential for the spherically symmetric,
self-gravitational droplet can be calculated using the formula
\cite{fliessbach2012allgemeine}
\begin{align}\label{m54}
\mathcal{U}^{\textsl{eff}}(r)=\frac{\exp(\Phi(r))}{2}\left(\xi+\frac{\ell^{2}}{r^{2}}\right),
\end{align}
where the terms $\ell$ and $\xi$ were already specified in Sec.
\textbf{IV}. By taking into account the fact that the effective
potential $\mathcal{U}^{\textsl{eff}}(r)\geq0$, we cannot compare
this value with the  Schwarzschild effective potential. When the
total mass $M$ associated with the self-gravitating droplet matches
the BH mass $M_{BH}$ at the galactic center, we consider the
rescaling as
\begin{align}\label{mv54}
\mathscr{L}=\frac{\ell}{r_{s}},~Q=\frac{q}{r_{s}},~\mathcal{H}=\frac{h}{r_{s}}, ~r^{\ast}=\frac{r}{r_{s}}, ~\textmd{with}~ r_{s}=2M_{BH},
\end{align}
then the values of the functions $B(r)$ and $\Phi(r)$ transformed as
\begin{align}\label{m55}
B(r^{\ast})=1+\frac{Q^{2}}{r^{\ast2}}-\frac{r^{\ast2}}{\mathcal{H}^{3}}\exp\left[-
\left(\frac{r^{\ast}}{\mathcal{H}}\right)^{\frac{1}{\beta}}\right]f(r^{\ast}),
\quad \textmd{where} \quad
f(r^{\ast})=\sum^{\infty}_{n=0}\frac{\left(\frac{r^{\ast}}{\mathcal{H}}\right)^{\frac{n}{\beta}}}{\Gamma(3\beta+n+1)}.
\end{align}
\begin{align}\label{m56}
\Phi(r^{\ast})=\frac{1}{\mathcal{H}^{3}}\int\frac{r^{\ast}\exp\left[-
\left(\frac{r^{\ast}}{\mathcal{H}}\right)^{\frac{1}{\beta}}\right]}{B(r^{\ast})}\left(\frac{2}{\beta\Gamma(
3\beta)}+2\frac{Q^{2}}{r^{\ast3}}-3f(r^{\ast})\right)dr^{\ast}.
\end{align}

The Einasto density-based self-gravitational model subject to the
non-local EoS provided in this section is not captivating for
reproducing the galactic motion of S-stars but can be important for
the modeling of DM mass coupled with anisotropic fluid allows
stabilized trajectories in case of massive particles. Although, this
type of self-gravitational droplet is not appropriate for the
modeling of central galactic cosmological structures but can be
modeled in other galactic areas. Finally, we calculate the values of
the physical variables, i.e., the fluid's pressure and the energy
density at the center of the gravitational droplet, and compare
their values with degenerate matter. Particularly, we take the
example of a stellar structure such as a neutron star with
degenerate pressure of the order $10^{31}-10^{34}~\textmd{Pa}$ and
the usual energy density $10^{17}~\textmd{kg/m}^{3}$, however in our
case, we assume $M=10M_{\odot}$. From the expressions \eqref{s9} and
\eqref{m37}, we have
\begin{align}\label{m57}
p_{r}(0)=\frac{c^{2}}{3}\rho(0), \quad \textmd{with} \quad\rho(0)=\frac{M}{4\pi h^{3}\beta\Gamma(3\beta)}\leq\frac{3.387}{4\pi h^{3}}.
\end{align}
In the above relation, we have made use of the fact that the function
$1/\beta\Gamma(3\beta)$ exhibit absolute maximum at $\beta=0.1538$ for $\rho(0)$.

\section{Conclusion}

This paper examines the effects of the FDM cosmological model on BHs
in the presence of electric charge, using a particular density
profile proposed by Einasto. This investigation enables us
to describe the connection between BH physics and the DM within the
background of a spherically symmetric matter distribution. This
intriguing correlation between cosmic components interacts with
gravitational fields on vast scales and influences the dynamics of
galaxies and clusters, leading to new insights.  We show that
coupling the Einasto density model with an anisotropic stress energy
tensor makes it possible to formulate various black hole solutions with specific choices of EoSs. We observed that assuming the EoS $\rho=-p_{r}$ leads to the formation of a BH solution, which
depends on the values of rescaled mass parameter $\omega$ or
spherically symmetric self-gravitational droplet. We could refer to
such an entity as a fuzzy BH if there is no horizon. Although the
concept of connecting the galactic centers to the DM is analogous,
these DM objects differ from the DM clusters created in
\cite{boshkayev2019model} their natural composition. Additionally,
we have calculated the expression of Hawking temperature for the
Einasto-inspired BH solutions within the framework of charged
Einstein's gravitational model. For both scenarios, i.e., for fuzzy
BH or fuzzy droplet, it is feasible to derive the effective
potential that controls the equations of motion. This potential
governs the orbits similarly as observed in the case of usual
galactic BH. It is noted that the assumption of a non-local EoS
rather than a local one gives rise to a self-gravitational droplet.
However, in this case, the negative value of $p_{r}$ appears to be
inevitable.

Additionally, this study reveals the existence of an unstable orbit
for massive particles with low angular momenta, as well as the
formation of bound states. Our analysis also showed that the Einasto
density model extends the extension of the Gaussian density profile.
Consequently, by considering a more generic formalism the problems
relating to the exotic spacetime configuration or perturbative
microscopic BH solutions \cite{batic2019perturbing} can be addressed
under analogous approaches. It is worth mentioning that our
conclusions are consistent with the results studied in
\cite{becerra2021hinting} but by adopting different methods.
Specifically, the numerical results presented in \cite{park2015no}
demonstrate the presence of DM in the supermassive BH, Sagittarius
$A^{\ast}$, composed of ``darkinos", as well as a description of the
BH through G2 gas clouds.

\vspace{0.3cm}


\section*{Acknowledgement}

The work by BA was supported by Researchers Supporting Project number: RSPD2024R650, King Saud University, Riyadh, Saudi Arabia. The work of KB was supported by the JSPS KAKENHI Grant Number 21K03547, 23KF0008 and 24KF0100.

\section*{Data Availability Statement}

This manuscript has no associated data or the data will not be deposited. [Authors comment: This manuscript contains no associated data.]

\vspace{0.3cm}

\renewcommand{\theequation}{A\arabic{equation}}
\setcounter{equation}{0}

\end{document}